# A versatile scanning acoustic platform

**N G Parker, P V Nelson and M J W Povey**

School of Food Science and Nutrition, University of Leeds, Leeds, LS2 9JT, United Kingdom

Email: n.g.parker@leeds.ac.uk, m.j.w.povey@leeds.ac.uk

**Abstract.** We present a versatile and highly configurable scanning acoustic platform. This platform, comprising of a high frequency transducer, bespoke positioning system and temperature-regulated sample unit, enables the acoustic probing of materials over a wide range of length scales and with minimal thermal aberration. In its bare form the platform acts as a reflection-mode acoustic microscope, while optical capabilities are readily incorporated to extend its abilities to the acousto-optic domain. Here we illustrate the capabilities of the platform through its incarnation as an acoustic microscope. Operating at 55 MHz we demonstrate acoustic imaging with a lateral resolution of 25 microns. We outline its construction, calibration and capabilities as an acoustic microscope, and discuss its wider applications.

## 1. Introduction

Acoustic waves provide a powerful and versatile modality with which to probe a wide range of media. Conveyed through the elastic response of a material, these perturbations can access deep into materials, including many that are optically opaque, and interrogate their internal elastic properties. These facets have been widely exploited in medical imaging [2], non-destructive testing of materials [3, 4] and industrial fluid characterisation [5, 6].

A key feature of acoustic waves is their scalability. Sound can be employed from global scales, e.g. in ocean and seismic tomography, down to the micro- and nano-scale. This latter regime is the realm of the scanning acoustic microscope (SAM), a precision device which uses tightly focussed ultrasound to map the acoustic contrast within a sample [7, 8]. The first SAM was demonstrated by Lemons and Quate in 1974 [9] with a resolution of approximately 10 microns. Following this the resolution was progressively reduced, through optical [10] and sub-optical [11] resolution, to the current record of 15 nanometres (150 Angström) [12]. The latter case was performed in a cryogenic environment of superfluid Helium. In the more convenient environment of water a resolution of below 200nm has been achieved at an operating frequency of 4.4 GHz and an ambient temperature of 60 °C [11].

Acoustic boundaries generate reflections whose time of flight details the surface and sub-surface structure of the sample. Furthermore, the amplitude and phase of the returning signal carries mechanical information which can reveal elastic moduli, compressibility, stress [13], adhesion properties, thermophysical properties and phonon transport [14]. Acoustic waves are highly complementary to optical methods. Take, for example, oil-in-water emulsions. While the oil-water interface generally possesses a small refractive index difference, there is a large acoustic contrast due to thermal scattering. This is valuable, for example, in studies of oil crystallization in emulsions where the acoustic contrast is very sensitive to the liquid-solid phase transition [15, 16]. For such reasons acoustic microscopy has exciting applications in determining the physical behaviour of a range of materials, including construction, biological and colloidal materials.

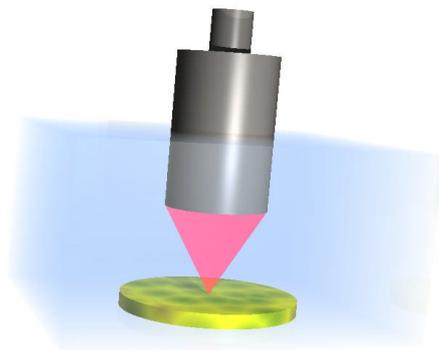

Figure 1: Schematic of the reflection-mode scanning acoustic microscope. A transducer unit generates and focuses an ultrasound pulse through a coupling fluid onto a sample. Reflected waves detected by the transducer unit reveal the acoustic properties of the sample.

Acoustic microscopes operate either in transmission or reflection mode. The former requires separate and opposing emitting and receiving transducers, and is limited to transmissive samples. In the latter case a single transducer acts as both transmitter and receiver of reflections from the sample. Due to its ease of arrangement and alignment, reflection-mode is most commonly used and will be considered here. The operating principle is illustrated schematically in figure 1. Sound waves generated by a piezoelectric element within the transducer unit are conveyed along a buffer rod to a spherical lens. The lens focuses the waves onto the sample through a coupling fluid, and the subsequent reflections are detected. (Note that spherical transducer elements, which negate the need for a spherical lens, have also been employed [17]). The transducer unit or sample is then scanned in space to form a spatial image of the acoustic contrast of the sample.

The coupling of optical and acoustic waves opens up further possibilities for gathering optical and functional information through photo acoustics [18] and ultrasound-modulated optical tomography (USMOT) [19-22]. In the latter case, a precision acoustic beam is employed to 'tag' light travelling through a turbid medium. The tagging occurs through the acoustic modulation of such optical properties as refractive index and the concentration of scatterers and absorbers. This technique enables the scattered light to be spatially localised within the sample to the depth and resolution of the acoustic beam. Importantly, this has enabled optical (including fluorescence) imaging in biological samples to depths that are well beyond the conventional scattering limit. While early work on USMOT employed low acoustic frequencies of around 1 MHz [19-22], recent work is moving towards the higher frequencies found in acoustic microscopes. For example, working at 75 MHz, optical imaging in tissue phantoms at a resolution of around 30 microns and to a depth of over 2mm has recently been achieved through this method [23]. This micro scale resolution marks the advent of ultrasound-modulated optical *microscopy*. For the purposes of this work it should be noted that the simplest geometry for performing ultrasound-modulated optical imaging consists of fixed, wide-field laser illumination while a tightly focussed acoustic beam scans the sample.

We have constructed a versatile scanning acoustic platform (VSAP) comprising of a high frequency transducer, bespoke positioning system and tightly regulated sample unit. In its bare form it operates as an acoustic microscope. However, it also incorporates optical capabilities enabling it to perform acousto-optic imaging. Our platform opens the possibility of obtaining both acoustic and optical contrast in thick, turbid tissues and to high resolution. The bespoke positioning system enables a large range of motion and micron resolution. Further, the tight temperature regulation within our sample unit dramatically reduces thermal aberrations, as well as supporting temperature-sensitive samples, such as biological tissue and emulsions. We illustrate our platform through its incarnation as a reflection-mode scanning acoustic microscope. Operating at a frequency of 55 MHz (-6 dB bandwidth of 20 MHz) our VSAP demonstrates a resolution of 25 microns. We present in detail the design, construction and calibration of the VSAP and its function as an acoustic microscope. We begin in Section 2 by introducing the background acoustics pertinent to understanding acoustic microscopy. In Section 3 we detail the construction of the acoustic platform, discussing the main components in turn. In Section 4 we discuss the operation and calibration of these components. In Section 5 we present some example results of our acoustic microscope. Finally in Section 6 we discuss extensions of this platform to acousto-optics and summarise our work.

## 2. Underlying acoustics

We here introduce the acoustical theory that underpins the design and operation of our acoustic microscope.

### 2.1. Acoustic waves

Acoustic waves are composed of elastic vibrations between adjacent particles in a material. On a macroscopic scale this forms a travelling pressure wave. In bulk materials acoustic waves are either longitudinal or shear depending on whether the particle motion is parallel or perpendicular to the wave direction, respectively. Shear waves are rapidly dissipated in elastic fluids and so we will consider longitudinal waves throughout this work unless otherwise stated. We denote the longitudinal speed of sound by $v$. A perfectly elastic and homogeneous medium supports plane pressure waves of the form,

$$p(x,t) = p_0 \exp[i(kx - 2\pi ft)] \qquad (1)$$

Here $p_0$ is the pressure amplitude, $f$ is the wave frequency and $k=2\pi/\lambda$ is the wave number, where $\lambda=v/f$ is the wavelength. Note that the behaviour is more complex at a boundary of the material, e.g., the creation of Rayleigh waves and shear waves at a solid-fluid interface, together with thermo-acoustic effects.

## 2.2. Acoustic contrast

Reflection-mode acoustic microscopy detects sound waves that have been reflected and back-scattered. These phenomena both arise from boundaries in the elastic distribution, which is usually parameterised through the characteristic acoustic impedance $Z=\rho v$, where $\rho$ is the density. Broadly speaking, the interaction is a reflection when the length scale of the boundary/bounded object $l$ is much greater than the sound wavelength $\lambda$.. Meanwhile, when $l \leq \lambda$, the feature acts as an inhomogeneity and scatters the sound. Scattering itself has a spectrum of behaviour, ranging from the mid-frequency regime ($l\sim\lambda$), where the scattering is sensitive to shape and size resonances [24], up to the far limit of Rayleigh scattering where the scattering becomes insensitive to shape ($l>>\lambda$) [25].

Consider plane sound waves in medium I ($Z_I$, $v_I$ and $\rho_I$) at normal incidence to an interface with medium II ($Z_{II}$, $v_{II}$ and $\rho_{II}$). The reflection coefficient $R$, the ratio of the reflected pressure amplitude $p_R$ to the incidence pressure amplitude $p_I$, is [3],

$$R = \frac{Z_{II} - Z_I}{Z_{II} + Z_I}. \qquad (2)$$

Equation (2) provides important intuition; reflection increases with the size of the impedance mismatch. However, strictly, $R$ is a function of incident angle $\theta$ and can change markedly with $\theta$, e.g., due to critical angles at which complete internal reflection or surface wave generation can occur. Indeed, the tightly focussed beams employed in SAMs can include incident angles of up to 60°. The true reflected signal thus requires a non-trivial integration of the reflection function over this angular distribution [26].

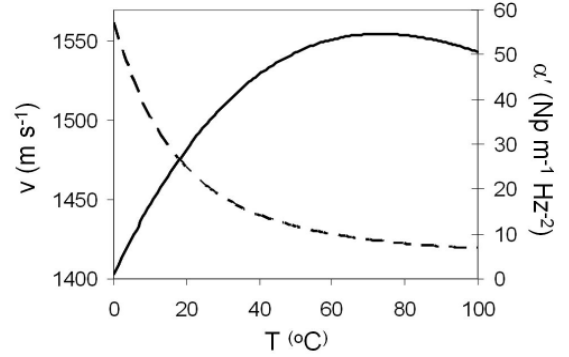

Figure 2: Speed of sound $v$ and attenuation coefficient $\alpha'$ in distilled water as a function of temperature. Data is taken from reference [1].

## 2.3. Focussing aberrations

In chromatic aberration different frequency components in the beam are refracted by the lens to differing degree, leading to a spectral spread in focal position. However, acoustic microscopes are usually sufficiently dominated by a single frequency that chromatic aberration is not a significant effect [8].

In an ideal spherical lens under monochromatic insonification, all paraxial incoming rays will be refracted to a common point at a focal distance $F$. In reality, rays at different radii from the lens axis have different focal distances, causing spherical aberration. Third-order theory reveals that the deviation of focal distance scales as $(v_{cf}/v_l)^2$ [8], where $v_{cf}$ and $v_l$ are the speeds of sound in the coupling fluid and lens. Since the speed mismatch is typically large (e.g. for a quartz-water boundary $v_{cf}/v_l\sim0.25$) this deviation can often be negligible in acoustic microscopy.

At ultra-high frequency and for wide-aperture lenses aberrations can become considerable, e.g. due to the differential attenuation experienced by the different beam path lengths. These complex effects are now well-understood [27].

## 2.4. Resolution

As described above, lens aberrations in acoustic microscopes can often be sufficiently small that the imaging resolution is close to the diffraction-limit. Consider a plane wave diffracting through a circular aperture of diameter $D$ (the lens). According to far-field wave theory

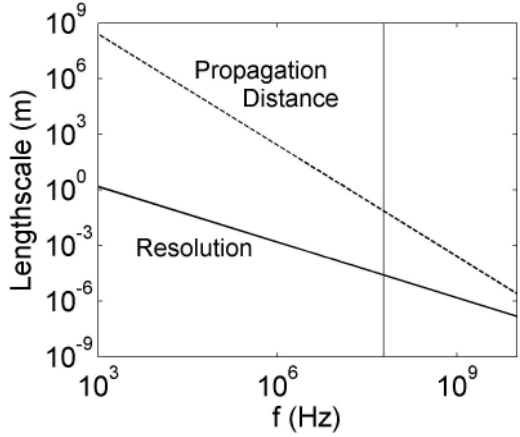

Figure 3: Resolution [Equation (4) assuming $F \approx D$] and propagation distance [Equation (6)] of a SAM in distilled water at $30^\circ$C ($v$=1509 ms$^{-1}$) as a function of frequency $f$. The vertical line indicates our operating frequency.

the pressure field $p(r)$ in the focal plane varies with radial position $r$ according to,

$$p(r) = p_0 \frac{J_1(\pi Dr/F\lambda)}{\pi Dr/F\lambda}, \quad (3)$$

where $J_1(x)$ is the Bessel function of the first kind and $F$ the focal length. The first node of this pressure distribution lies at position $d$=1.22$F\lambda$/D. According to the well-known Rayleigh criterion two objects are just resolvable when their separation equals this value. For the emitter/receiver system considered here the result become slightly modified to become [28],

$$d = 1.02 \frac{F\lambda}{D}. \quad (4)$$

NB the definition of resolution is somewhat arbitrary and best determined experimentally.

For a given lens (fixed $F$ and $D$), resolution is enhanced at greater frequencies. Figure 3 shows the resolution of the focussed sound beam in water at $30^\circ$C (assuming F≈D, which is typically valid). At kHz frequencies the resolution is of order a metre, while at GHz frequencies the resolution is less than a micron. The resolution can also be improved by a factor of √2 by operating in the nonlinear regime [29].

### 2.5. Thermal aberrations

Distances are inferred from the time of flight of the returning echoes and the speed of sound of the coupling fluid. However the speed of sound varies with temperature at up to 3 ms$^{-1}$ per $^\circ$C in water (see figure 2). This can introduce considerable thermal aberration in the image and so it is essential for temperature stabilisation. Moreover, the effect of temperature variations can be reduced by operating at temperatures with minimal gradient, e.g. around $70^\circ$C for water.

### 2.6. Attenuation

The plane wave solution (1) is an ideal solution of the propagation problem. Real fluids possess viscosity and finite thermal conduction which dampen the beam during propagation and result in an attenuated form,

$$p(x) = p_0 \exp[-\alpha x]\exp[i(kx - 2\pi ft)]. \quad (5)$$

For almost all fluids at the frequencies of interest the attenuation coefficient $\alpha$ has a quadratic dependence on frequency and can be expressed as $\alpha = \alpha' f^2$.

After some propagation distance the beam will be too weak to be detected. We define a propagation distance based on an arbitrary 100-fold (40 dB) decrease in beam amplitude, which is given by,

$$L_{0.01} = -\frac{\ln(0.01)}{\alpha' f^2}. \quad (6)$$

Figure 3 presents $L_{0.01}$ as a function of frequency for water at $30^\circ$C ($\alpha'$=18×10$^{-15}$ Np m$^{-1}$ Hz$^{-2}$). The propagation distance decreases more rapidly than the resolution and at some frequency becomes restrictively small. For example, at 1 GHz in water at $30^\circ$C the maximum propagation distance is limited to around 50µm. The presence and nature of any intervening interfaces will further reduce the signal amplitude and maximum propagation distance.

The attenuation coefficient changes with temperature and this can be exploited to generate conditions with reduced beam loss. For water, shown in figure 2 (dashed line), it is preferable to operate at raised temperatures, e.g. $\alpha'$ is halved from 20 $^0$C to 50 $^0$C.

## 3. Construction

An image of our VSAP is presented in figure 4. We will discuss the key components in turn.

### 3.1. Transducer unit

The transducer unit is a commercial high frequency focussed unit (Panametrics V3534) with a quoted fundamental frequency $f$=100 MHz. A fused quartz delay rod serves to temporally separate the reverberations in the unit. A spherical lens of diameter $D$ = 6 mm ground into the end of the delay rod focuses the pulse at a quoted distance $F$ = 5 mm. Due to the large reflection at the fluid-lens interface a quarter-wave layer is employed to enhance transmission. NB the operating frequency and focal position will be experimentally examined in Section 4.1.

### 3.2. External controls

A pulse-receiver unit (UTEX-320) generates a square wave pulse of 50ns duration and 300V amplitude to excite the transducer, and receive the returning signal. The signal is digitised and averaged through a digital oscilloscope. A computer is employed to synchronise the data recording with the position system and visualise the signal.

### 3.3. Sample unit

An integrated sample unit, composed of aluminium, provides both an inner well and a surrounding temperature bath, as illustrated in figure 5. The inner well contains the sample and coupling fluid (here Millipore water), and the transducer is immersed in the coupling fluid from above. Water from an external temperature bath (Haake DC50 circulator and Haake B5 bath) circulates through the outer annular well to regulate the temperature of the sample unit, with an embedded thermometer providing feedback to the external unit. Thermal insulation is provided by a layer of foam at the top of the annular well and flexible sealing film stretched over the top of the whole unit (through which the transducer penetrates).

### 3.4. Optical access

High quality glass windows of 3 cm diameter are embedded into opposing walls of the sample unit (not illustrated in figure 5). This allows optical access into the sample unit. Here we exploit this to perform simultaneous optical imaging of the sample. We employ a CCD camera mounted on a Leica Monozoom® 7 parfocal lens, which offers a field of view ranging from 20mm to 2mm and an optical resolution of down to 10 microns. Importantly, the optical access of our instrument can enable a laser system to be incorporated so as to perform acousto-optics, notably ultrasound-modulated optical tomography.

### 3.5. Positioning system

To provide versatility to image over ranging length scales we require a positioning system that combines high spatial precision with a large and configurable range of motion. We have constructed a bespoke device that offers a spatial resolution of the order of a micron and a range of several cms, all within a single positioning system. Our positioning system, constructed of arms and rotational joints, is based on arcular motion rather than the more conventional *xyz* motion. This design offers a simplicity that minimises the need for commercial parts and enhances the economy of the system. The transducer is connected to the bottom of a vertical arm and is itself moved (as opposed to moving the sample bath). This feature provides an open geometry which facilitates ease-of-access to the sample bath and readily accommodates the introduction of different transducers. These benefits come at the expense of compactness, robustness and the ultimate positioning resolution possible (e.g. compared to *xyz* positioning systems available). However, it is well-suited to the needs of our instrument.

Each positioning axis is driven by a stepper motor (Parker Powermax II) and timing belt. The *x*, *y* and *z* axles meet at an origin, to which the transducer unit is connected. The translation of each axle is restricted by two motion limiters, with a maximum translation range of 2 cm.

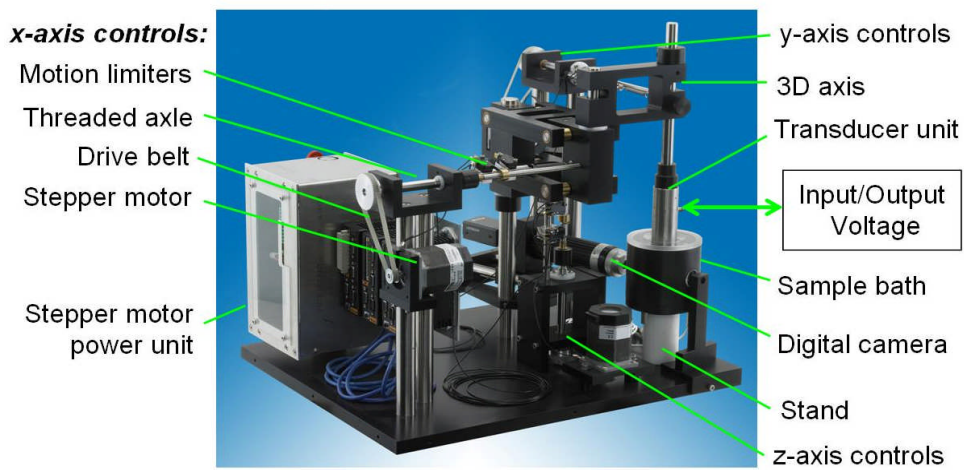

Figure 4: Image of the SAM with key components highlighted. The sample unit is shown in further detail in Figure 5.

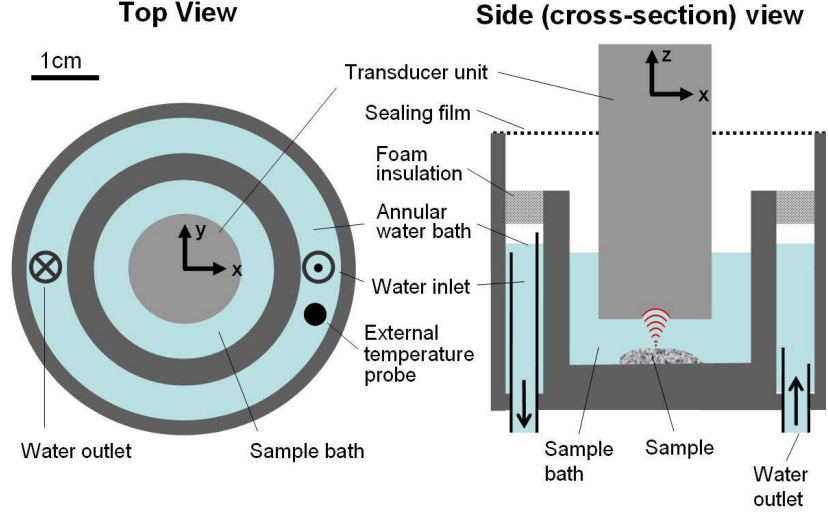

Figure 5: Schematic views of the sample unit from the top and side (cross-section). Blue/light grey denotes water, intermediate grey denotes the transducer unit, and dark grey denotes the aluminium housing. In the top view the foam insulation and jacket are not shown.

The axles are of fixed length and so the resulting motion is constrained to spherical surfaces (angular motion is made possible by ball-and-socket joints at the edge of each axle). Consider the *x-y* plane. Under translation of the *x*-axle by a distance *x* the transducer will trace out an arc of angular size $\theta \sim x/R$, where $R$ is the axle length. By geometrical arguments the projected distance translated along the *x*-axis is $R\sin\theta$, such that the position deviation is,

$$\Delta x = x - R\sin\theta. \qquad (7)$$

The arcular motion also leads to the growth of a displacement in the *y*-direction given by,

$$\Delta y = R(1 - \cos\theta). \qquad (8)$$

These deviations will be the same for the *x-z* and *y-z* planes. For our system $R$=16cm and so a typical range of *x*=1mm leads to deviations of $\Delta x$=7 nm and $\Delta y$=3 μm, which are small enough to be neglected. The precision and accuracy of the positioning will be measured in Section 4.3.

## 4. Operation and Calibration

### 4.1. Transducer/Beam characteristics

The properties of a high frequency transducer can often deviate from its ideal, quoted properties. As such it is essential to characterise the transducer experimentally. Further details can be found in the works of Shiloh *et al.* [30] and Lee *et al.* [31, 32] who detail the characterisation of transducers of similar construction and frequency.

*4.1.1. Transducer Electrical Impedance.* As an electro-mechanical device one can characterise the transducer electrically. A plot of the electrical impedance of the transducer as a function of frequency is shown in Figure 6. This is obtained by connecting the transducer to a network analyser (Agilent, E5062A, range 300kHz-3GHz). Impedance contributions from the connecting cables are observed to be very small in comparison. During measurement the transducer is immersed in water to simulate its working conditions. The impedance response resembles that of an LC resonator circuit, with a clear resonance at ~50 MHz and anti-resonance at ~95 MHz. The UTEX pulser-receiver is a low impedance system of approximately 50 Ω, hence maximum power transmission into the transducer can be expected when the electromechanical impedance of the transducers is of this order, i.e. at around 60 MHz and 100 MHz, the two main output frequencies of the transducer.

*4.1.2. Transducer emission.* Insight into the transducer emission can be obtained from the first acoustic reverberation within the transducer. This pulse is shown by the grey line in Figure 7(a). Its Fourier spectrum, shown by the grey line in Figure 7(b), features a primary component at 61 MHz and a secondary component at 100 MHz. By careful optimization of the circuitry and excitation pulse we could preferentially excite the 100 MHz component so as to ensure maximal imaging resolution with optimal signal-to-noise. However, having a range of well-defined frequencies present provides versatility to vary the operating frequency

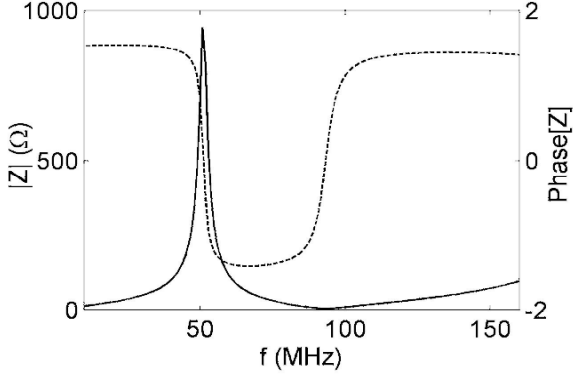

Figure 6: Complex electrical impedance (solid line) and phase (dotted line) as a function of frequency for our transducer.

as desired, e.g. for the examination of frequency resonances or attenuation spectra.

*4.1.3. Signal Properties.* We insonify a flat surface of polytetrafluoroethylene (PTFE) placed in water at 30 $^0$C. PTFE is employed because its reflection function is approximately flat over the range of ray angles generated by our transducer [8] and will not generate surface acoustic waves (which would complicate our results). The signal at focus is shown by the black line in Figure 7(a). The Fourier spectrum of the focal signal is shown by the black line in Figure 7(b). The primary frequency component is now at 55 MHz and the secondary component is at 83 MHz. We can readily explain and estimate this frequency shift by considering a generic pulse with Gaussian frequency distribution,

$$V(z=0) = V_0 \exp\left[-\frac{(f-f_0)^2}{2\sigma^2}\right] \qquad (9)$$

where $V_0$ is the initial peak voltage of the pulse, $f_0$ is the initial centre frequency and $\sigma$ is the width of the frequency distribution. If the transducer-surface distance $z$ is non-zero, the returning voltage will be attenuated as per Eq. (5) leading to a modified voltage,

$$V(z) = V_0 \exp\left[-\frac{(f-f_0)^2}{2\sigma^2} - 2\alpha' f^2 z\right]. \quad (10)$$

The peak frequency will occur when the exponent is minimal, that is, when,

$$f_0(z) = \frac{f_0}{1+4\alpha' z\sigma^2} \approx f_0\left(1-4\alpha' z\sigma^2\right), \qquad (11)$$

where we have used the Taylor series expansion to give an approximate form valid when $4\alpha' z\sigma^2 \ll 1$. Acoustic attenuation thus causes the peak frequency to shift to lower frequencies during propagation. Taking $\alpha' = 18 \times 10^{-15}$ Np m$^{-1}$ Hz$^{-2}$, $\sigma = 20$ MHz and $z = 5.8$ mm (see later) then Eq. (11) predicts that the 100 MHz peak will shift to 83 MHz and the 60 MHz will shift to 51 MHz. This is consistent with the frequency shifts observed in Figure 7(b).

Note that we will henceforth consider the operating frequency of our VSAP to be 55 MHz. The -6dB bandwidth is measured to be 21 MHz.

*4.1.4. Detection properties*

Our signal is detected and undergoes analogue-to-digital conversion by an oscilloscope (Lecroy Waverunner Xi-64). We typically average over 100 sweeps to reduce random noise. The dynamic range of the oscilloscope data is measured to be 90.5 dB. However, spurious signals of the order of 100µV limit the lower end of this range. Hence for typical voltage amplitude of 100mV we obtain a signal-to-noise ratio of 60 dB.

*4.1.5. Focal position*

The true focal position can be conveniently located by considering how the pulse amplitude varies with the transducer-surface separation, as demonstrated by Lee and Bond [32]. In Figure 7(c) we show the so-called $V(z)$ curve from the PTFE surface. It is sharply peaked in the focal region, where the reflecting waves are in phase and give maximal signal. The voltage decreases away from focus due to the dephasing of the focussed waves and, for surfaces beyond the focal point, the geometric fact that a proportion of the reflecting rays fall outside the lens [32]. We can thus estimate the focal point as the point at which the peak voltage is a maximum, giving $z_F$=5.791 mm. This definition is crude, since focal point will vary with frequency due to chromatic lens aberrations and frequency-dependent attenuation in the medium. More precisely, the focal point should be measured at the operating frequency [32]. The magnitude of the 55 MHz pulse component is shown by the grey dashed line in Figure 7(c). It shows a small deviation from the earlier value, giving an improved focal position of $z_F$=5.795 mm.

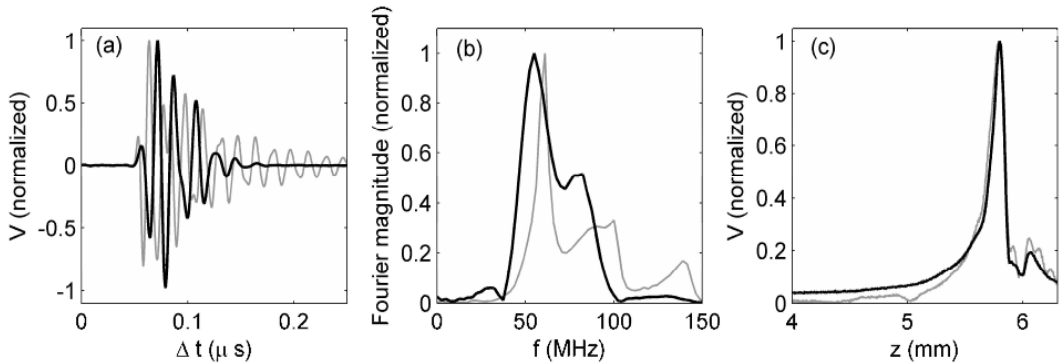

Figure 7: (a) Signals corresponding to the first reverberation in the transducer unit (grey) and the reflection from a PTFE surface (black) located at the focal point in water at 30$^0$C. The amplitude of the focal pulse is 62 mV. (b) Magnitude of the Fourier spectrum of the pulses in (a). (c) Peak voltage of the raw pulse (black line) and the 55 MHz component (grey dashed line) as the transducer-sample distance $z$ is varied.

*4.1.6. Resolution.*

The lateral resolution is estimated from Equation (5) to be $d$=23 μm (assuming water at 30 $^0$C and a frequency of 55 MHz). The lateral resolution/beam spot size can be determined experimentally by considering the beam interaction with sharp objects such as needles, edges or line objects [30]. Of these, the edge response is particularly convenient and accurate, and has been employed successfully elsewhere [31, 32, 30]. Glass provides a sharp edge and here we employ a glass microscope slide. The upper surface of the slide is located in the focal plane and the beam is scanned across the edge (in the $x$-direction). The voltage signal in $x$-t space (figure 8 (b)) clearly shows a step in the echo time due to presence of the edge. The voltage profile in the focal plane, which is presented in figure 8 (c), shows a smooth transition to zero due to finite beam width. An excellent description of the theoretical edge response for different ultrasonic imaging systems is presented in reference [31]. For a confocal system, the received voltage at focus is given by,

$$V(x, z = F) = A \left| \frac{J_1(\pi Dx / F\lambda)}{\pi Dx / F\lambda} \right|^2 * H(-x_s). \quad (12)$$

Here $H(-x_s)$ is the Heaviside step function ($H$=1 for $x<x_s$ and $H$=0 for $x>x_s$) which models the step and $A$ is a normalisation factor that incorporates the electro-mechanical response of the system. The Bessel function $J_1$ was defined earlier in Eq. (3) and is squared here due to the compounded effects of emission and receiving. The beam width is determined by fitting Equation (9) to the measured $V(x,z=F)$ data. For a beam width of 25 μm we get an excellent agreement between the experimental data (solid line) and theoretical fit (dashed line). This width is also in very good agreement with the estimated value of 23 μm (see above). Note that by differentiating the edge response (13) we arrive at the line spread function and by Fourier transforming this result we can also derive the modulation transfer function [30].

The axial resolution is set by the pulse length. The -6 dB width of the focal pulse in Figure 7(a) is 25.6ns, corresponding to a distance in 30 $^0$C water of 39 μm. Meanwhile the depth of focus is given by the expression [27],

$$F_z = \frac{8F^2 v}{D^2 f + 2Fv}. \quad (13)$$

For water at 30$^o$C this gives $F_z$ = 200 μm.

*4.1.7. Point Spread Function.* The spatial distribution of the beam can also be examined through the point spread function (PSF). This is the signal returned by a point object as a function of the lateral distance between the lens and object. This approach has been considered elsewhere to characterise an ultrasound beam, for example, in references [30, 33, 34]. To make a quantitative measure of beam resolution the needle tip must be significantly smaller than the beam resolution. Since it is difficult to obtain needles this small for our interests, we are limited here to obtaining qualitative information about the beam profile and spreading. We image the tip of a wire which is 80μm in diameter. At a given sample distance we scan laterally across the object. The position of the sample is expressed relative to the focal point by the parameter *Z*, with *Z*<0 corresponding to the sample being in front of the focal point. The demodulated signal detected as a function of lateral displacement is presented in figure 9. At the focal point the PSF, and therefore the beam, is at its narrowest, and is single peaked. NB the PSF at the focal point is limited by the size of the pin tip, and so does not reach the true minimal value. The lateral extent of the PSF/beam increases as we move away from the focal point. Lobes form due to diffraction effects and the PSF develops a curvature due to the lens curvature. We have additionally calculated the PSF theoretically using the Field II simulation [33, 34], with the results presented in the right-hand column of figure 9. This method assumes linear acoustics and derives the PSF by summing the spatial impulse response from small segments of the concave lens, for which an analytic form exists [35, 36]. The PSF is then the convolution of the spatial impulse response with the excitation function, for which we assume some approximate form. Our experimental results are in good agreement with the theoretical predictions, including the beam width, curvature and presence of lobes. Deviations arise from a number of sources, including irregularities in the pin tip and lens apodization.

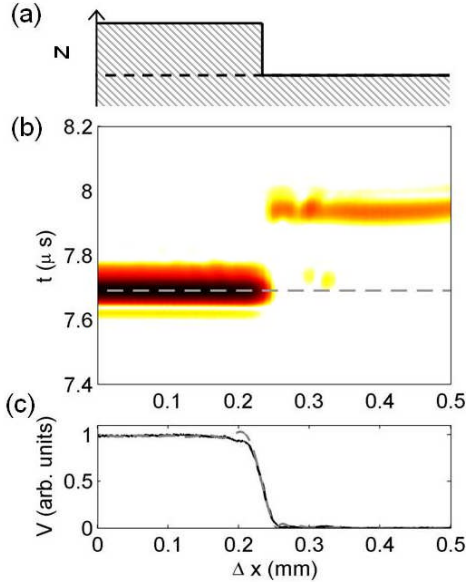

Figure 8: (a) A glass slide placed on another creates a sharp edge. (b) Voltage detected as a function of time and lateral position across the edge. The dashed line denotes the focal plane. (c) Received voltage in the focal plane, and the theoretical prediction (dashed line) of Eq. (12).

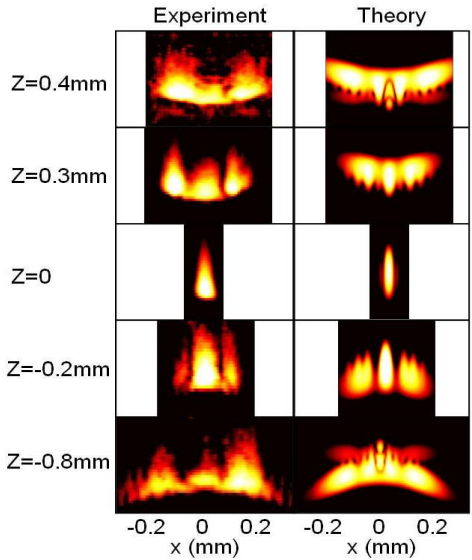

Figure 9: Point spread function as measured experimentally (left column) using a pin point and determined theoretically (right column) using the Field II simulation. The *y*-axis of each image represents the timescale of the returning signal.

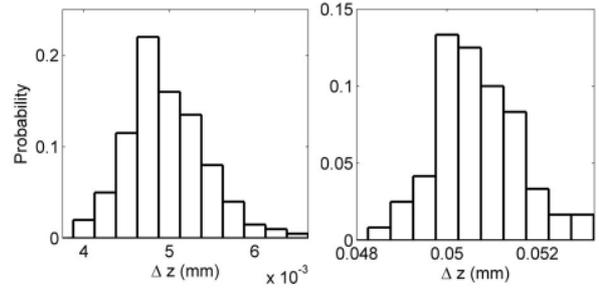

Figure 10: Probability distribution of the distance traversed by the positioning system as determined by time-of-flight measurements, for expected distances of 5μm (left) and 50μm (right).

*4.1.8. Angular reflection.* Another important consideration in beam detection is that of angular reflection from a tilted surface. For a surface which is normal to the incident beam the waves are reflected back to the transducer. However, tilted surfaces result in angular reflection which reduces the proportion of the reflected energy that falls back on the transducer lens. In this manner the surface topography modulates the returning acoustic amplitude. From simple geometrical arguments we expect that the received signal becomes negligible when the surface is tilted away from the horizontal by around $20^0$ and this is consistent with that observed experimentally.

## 4.2. Temperature bath

It is essential for a uniform ambient temperature between the transducer and the sample to prevent the thermal image aberrations discussed

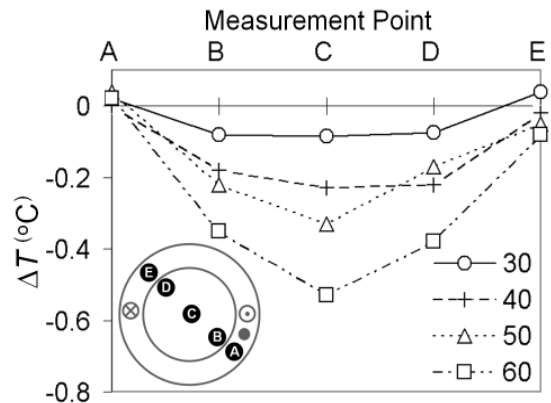

Figure 10: Difference between the actual temperature and the set temperature at five positions A-E across the sample unit, as indicated in the inset. The set temperatures are specified in the legend.

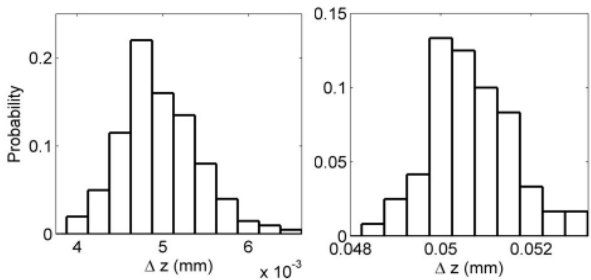

Figure 11: Probability distribution of the distance traversed by the positioning system as determined by time-of-flight measurements, for expected distances of 5μm (left) and 50μm (right).

in section 2.5. We have mapped the temperature distribution in the sample unit using a four wire platinum resistance thermometer (Hart Scientific 5612 probe), accurate to 0.01°C, traceable to ITS-90. After setting the desired temperature on the water circulator and allowing sufficient time to equilibrate we measured the actual temperature at five positions across the sample unit (see inset of figure 10). This was performed for set temperatures of 30, 40, 50 and 60 °C. Note that for access the transducer could not be put in place during these measurements. The deviation between the set temperature and the measured temperature $\Delta T$ is shown in figure 10. At point A, which is adjacent to the temperature probe regulating the bath, the measured temperature is in excellent agreement with the set temperature. Furthermore, the correspondence between the temperatures at points A and E show that the annular temperature bath is of approximately uniform temperature. The inner sample well is lower in temperature than the annular water bath due to imperfect thermal contact. Furthermore, within the sample well there exists a temperature distribution due to thermal losses and currents. At a set temperature of 30°C this temperature variation is approximately 0.01 °C. Taking into account the temperature dependence of the speed of sound (figure 2) this corresponds to a 0.003% variation in speed of sound, and therefore distance measurements, which is negligible. The temperature variation grows at larger ambient temperatures, being 0.2 °C at 60°C. However, the corresponding variation in speed of sound is approximately 0.01%, which is still small enough to be negligible for most purposes. The presence of the foam insulation and sealing film is crucial in preventing evaporation and heat losses, giving an order of magnitude improvement in the thermal variations; in their absence the temperature variations in the inner well were approximately

0.2°C at 30°C and 2°C at 60°. Hence, through careful control of temperature we have achieved at least a tenfold improvement in resolution. Further reduction in thermal variations could be made by improved thermal insulation and the inclusion of stirring to suppress thermal currents.

### 4.3. Positioning system

Each rotational increment on the stepper motor corresponds to a 50 nm translation of the axle. A large backlash occurs in the belt and its connection with the gears and is particularly evident during a change of direction. This is negated by performing saw tooth raster scanning where, after a full scan in one direction, the system is reversed back to its start point, ready for the next scan (rather than the conventional back-and-forth scanning). The start point can be conveniently defined by one of the motion limiters. Accuracy of motion is readily estimated in the $z$-direction by independently measuring the distance from the pulse time-of-flight. Figure 11 presents a probability distribution of the time-of-flight distance for steps of (a) 5 μm and (b) 50 μm. In both cased we observe a distribution centred on the desired distance. The width of the distribution is approximately 1μm in both cases. Since the error in the positioning is much less than the resolution, these errors have no significant effect.

## 5. Results

To illustrate the versatility and capabilities of the VSAP as an acoustic microscope we here present and briefly discuss some example images.

### 5.1. Onion skin

One of the first suggested applications of scanning acoustic microscopy was in the study of biological matter [37] due to its non-invasiveness and beneficial imaging contrast. This area has now grown with major applications being found in the study of bone and teeth, cells and soft tissue samples such as the eye and skin (see the reviews of references [8, 4, 38]). In figure 12(a) we present an acoustic image of onion skin. This is a C-scan image from the focal plane of the transducer, and was achieved by time-gating the signal over a 50ns gate about the focal plane. The focal plane was 30 microns beneath the uppermost cell surface, and so corresponds approximately to the mid-plane of the cell. Individual onion cells are clearly visible, even at modest imaging resolution. Excellent contrast of the macroscopic cellular structure is observed, with the membrane/wall exhibiting the strongest reflection. The image took approximately two hours to acquire. It is worth noting that the mechanical properties of cells and soft tissues can be extracted from acoustic microscopy using a variety of mathematical techniques [4, 8].

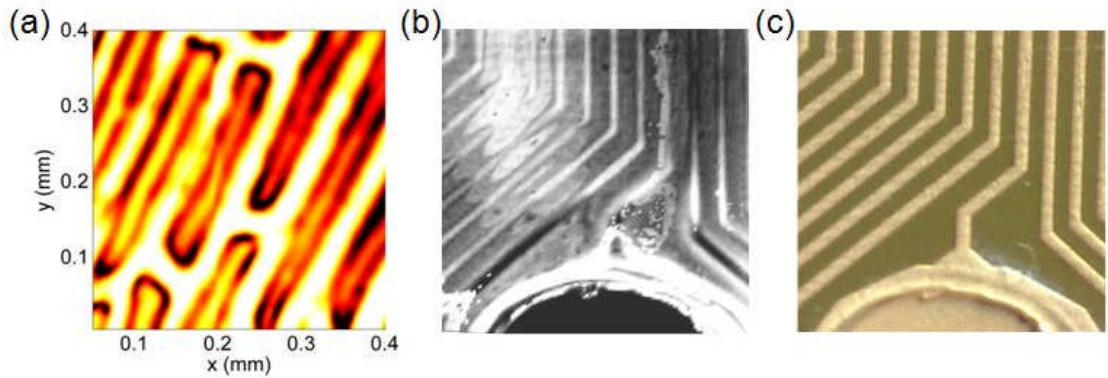

Figure 12: Images from the SAM. (a) Onion skin. (b) An integrated circuit (2×2mm) with the corresponding optical image presented in (c).

*5.2. Integrated circuits*

One of the largest applications of scanning acoustic microscopy has been in non-destructive testing of electronic circuitry [4]. Figure 12(b,c) presents a comparison between the acoustic and optical image of an integrated circuit. The acoustic image plots the amplitude of the reflection from the circuit surface. It is important to note that the image is taken through a plastic protective layer over the circuit. The oval fringes in the top left of figure 12(b) arise from the interference between successive echoes from the sample. It is clear that the acoustic image contains greater detail than the optical image, as noted elsewhere [39], likely to arise from acoustic-sensitive aberrations in the coating or sub-surface details.

# 6. Conclusions

We have presented a versatile scanning acoustic platform for performing acoustic and acousto-optical microscopy. The platform combines excellent performance, versatility for multi-modal imaging, and a simple design that promotes economy. We have demonstrated its capabilities through its ready incarnation as an acoustic microscope. The key components of the platform are:

i) *Focussed transducer*: Our current high frequency transducer operates at 55 MHz, generating a focal point that extends 25 microns laterally and 200 microns axially. Note that the platform is designed to readily accept different transducers as required.

ii) *Temperature-regulation*: A temperature-control system enables the sample temperature to be strictly regulated. For example, at 30°C the variation across a typical sample is less than 0.01°C, leading to a variation in the speed of sound of less than 0.005%. With such control, thermal aberrations are practically negligible. Note that by reducing thermal variations of speed of sound with careful temperature control we demonstrate that the resolution is improved by at least a factor of ten.

iii) *Positioning system*: Our bespoke positioning system enables a scanning precision of a micron and a range of motion of up to 2 cm. Furthermore, its simple design allows easy access and economy.

These components natively form a reflection-mode acoustic microscope, suitable for mapping the acoustic microstructure of a range of soft materials, such as colloids and living tissues such as plant and animal cells. The operation of the instrument has been demonstrated by acoustic imaging onion cells and integrated circuits. This incarnation of the VSAP is currently being developed to obtain quantitative images of the mechanical properties of cell walls.

Importantly, the platform is also designed to allow optical access. In our current acoustic microscope this is exploited to perform simultaneous optical imaging of the sample. However, a more exciting prospect is the ability to couple laser light with acoustics so as to perform hybrid imaging techniques. In particular, our platform is ideally suited to perform ultrasound-modulated optical tomography. A static laser beam can be introduced to illuminate the whole sample and the scanning acoustic beam employed to tag the light locally within the sample. Optical detection components would then detect the outgoing modulated light. The optical components can be readily incorporated into the system thanks to its spacious and accessible design. Indeed, we are currently working towards this, with the aim of performing ultrasound modulated optical microscopy to the same resolution as our acoustic microscope, namely 25 microns. Our ultimate goal is to apply this instrument to engineered tissue to provide synchronous acoustic and optical imaging of the mechanical, morphological and functional properties therein.

## Acknowledgements

We thank Mel Holmes for technical assistance and stimulating discussions.